\documentclass[prd,reprint,amsmath,amssymb,preprintnumbers,superscriptaddress,nofootinbib]{revtex4-1}

\usepackage{epsfig,epstopdf}
\usepackage{subfigure}
\usepackage{color}

\def\beq{\begin{equation}}
\def\eeq{\end{equation}}
\def\bea{\begin{eqnarray}}
\def\eea{\end{eqnarray}}
\def\nn{\nonumber}

\def\eps{\varepsilon}

\newcommand{\lsim}{
\mathrel{\hbox{\rlap{\hbox{\lower4pt\hbox{$\sim$}}}\hbox{$<$}}}}
\newcommand{\gsim}{
\mathrel{\hbox{\rlap{\hbox{\lower4pt\hbox{$\sim$}}}\hbox{$>$}}}}
\newcommand{\mat}[2][ccccc]{\left( \begin{array}{#1} #2\\ \end{array}\right)}

\newcommand{\dM}{\text{det}[M]}
\newcommand{\tM}{\text{tr}[M]}

\def\eps{\varepsilon}

\begin{document}

\preprint{CTPU-16-23}
\title{Gauge see-saw: A mechanism for a light gauge boson}

\author{Hye-Sung Lee}
\email{hlee@ibs.re.kr}
\affiliation{Center for Theoretical Physics of the Universe, Institute for Basic Science (IBS), Daejeon 34051, Korea}
\author{Min-Seok Seo}
\email{minseokseo@ibs.re.kr}
\affiliation{Center for Theoretical Physics of the Universe, Institute for Basic Science (IBS), Daejeon 34051, Korea}

\date{August, 2016}
\begin{abstract}
\noindent 
There has been rapidly growing interest in the past decade in a new gauge boson which is considerably lighter than the standard model $Z$ boson.
A well-known example of this kind is the so-called dark photon, and it is actively searched for in various experiments nowadays.
It would be puzzling to have a new gauge boson which is neither massless nor electroweak scale, but possesses a rather small yet nonzero mass.
We present a mechanism that can provide a light gauge boson as a result of a mass matrix diagonalization.
\end{abstract}
\maketitle

%%%%%%%%%%%%%%%%%%%%%%%%%%%%
\subsubsection{Introduction}
%%%%%%%%%%%%%%%%%%%%%%%%%%%%
It is amusing to observe that a square matrix of the equal size of entities
\bea
M = \mat{1 & 1 \\ 1 &1}
\label{eq:M1111}
\eea
results in the eigenvalues $\lambda = 0$ and $2$.
When the matrix is slightly tilted or misaligned from the original matrix, there will be a nonzero but tiny eigenvalue $\lambda \ll 1$.
There are even more general cases than the one presented in Eq.~\eqref{eq:M1111}.
In this letter, we will use this mechanism to rationalize a very light gauge boson.

A light gauge boson, sometimes called dark gauge boson (typically, MeV - GeV scale, but it can be even lighter) has been a popular subject to study after it was shown it could potentially address many puzzling observations such as the positron excess, the small scale problems around the galaxy, and the muon $g-2$ anomaly \cite{Essig:2013lka}.
If its lifetime is sufficiently long, the dark gauge boson itself can be a dark matter candidate \cite{Pospelov:2008jk,Nelson:2011sf,Arias:2012az}.

For such a light particle to survive all the experimental constraints, it should have a very small coupling.
A popular model is called the dark photon, because it couples only to the electromagnetic current like the photon when it is substantially lighter than the $Z$ boson of the standard model (SM) \cite{ArkaniHamed:2008qn}.
A dark $U(1)$ can mix with the hypercharge $U(1)_Y$ of the SM through a gauge kinetic mixing term $\frac{\eps}{2 \cos\theta_W} Z'_{\mu\nu} B^{\mu\nu}$, and couples to the SM particles through this mixing, which can be suppressed by the loops of some heavy fermions that have charges under both the dark $U(1)$ and the $U(1)_Y$ \cite{Holdom:1985ag}.

The smallness of the mass may be explained by taking the vacuum expectation value (vev) of a scalar, which is responsible for the dark $U(1)$ symmetry breaking, is also of very small scale.
Yet, it would be desirable to find a possible mechanism to obtain a very light gauge boson from the high scale (electroweak or UV scale) physics without introducing a new scale.
Some models that can address this using the supersymmetry framework can be found in Refs.~\cite{ArkaniHamed:2008qp,Cheung:2009qd,Kang:2010mh}.

In this letter, we will employ two massive gauge bosons of the same heavy mass scale and their large mixing to realize a similar mass matrix texture as Eq.~\eqref{eq:M1111} or even a more general form.
The mass matrix of this form can be realized with, for instance, Higgs mechanism or St\"uckelberg mechanism.
We shall call our mechanism {\em Gauge See-saw} as they rely on the mass matrix diagonalization like the neutrino see-saw to obtain a small mass for one particle while its partner remains in the heavy scale, although the mass matrix texture is very different from the typical (type-I) neutrino see-saw \cite{Minkowski:1977sc,seesaw}.

There are some relations between the properties of the two gauge bosons in our mechanism, and a discovery of one particle can help in searching for the other particle.
We will discuss some implications of the gauge see-saw later in this letter.

%%%%%%%%%%%%%%%%%%%%%%%%%%%%
\subsubsection{Gauge See-saw}
%%%%%%%%%%%%%%%%%%%%%%%%%%%%
For a $2 \times 2$ gauge boson mass-squared matrix
\bea
M = \mat{a & b \\ b &d} ,
\label{eq:Mmatrix}
\eea
the eigenvalues (physical mass-squared values) are given by
\bea
\lambda = \frac{1}{2} \left( \tM \pm \sqrt{\tM^2 - 4\,\dM} \right)
\eea
where
\bea
\dM &=& ad - b^2 \, , \label{eq:dM} \\
\tM &=& a + d \, .
\eea
%\bea
%\dM &=& \lambda_1 \lambda_2 , \\
%\tM & =& \lambda_1 + \lambda_2 .
%\eea

The diagonal mass-squared terms ($a$, $d$) are always positive-definite.
While the off-diagonal mixing term ($b$) can be negative, it appears only in squared ($b^2$) in Eq.~\eqref{eq:dM}.
Thus, $\dM$ always contains a destructive sum, possibly resulting in a significant suppression from the original scales, while $\tM$ always has a constructive sum.
When all elements ($a$, $b$, $d$) are at the same scale, $\tM$ should remain at the original scale, while the $\dM$ can be orders of magnitude smaller in principle.

We define a mass alignment parameter $r$ as
\bea
r \equiv \frac{\dM}{\tM^2} \, .
\eea
The gauge see-saw can be achieved for $r \ll 1$, under which the physical masses of two gauge bosons ($Z_L$, $Z_H$) can be well approximated as
\bea
m_{Z_L}^2 \simeq \frac{\dM}{\tM} \, , \quad m_{Z_H}^2 \simeq \tM \, ,
\eea
and the mass alignment parameter itself clearly shows the disparate mass scales as
\bea
r \simeq \frac{m_{Z_L}^2}{m_{Z_H}^2} \ll 1 \, . \label{eq:rapprox}
\eea
A GeV-TeV level mass hierarchy would require $r \approx 10^{-6}$.
In the perfect mass alignment case ($r = 0$), $Z_L$ becomes massless.\footnote{In this limit, there are similar aspects with Ref.~\cite{Shiu:2013wxa}, in which a certain kind of mass matrix was exploited to realize the massless gauge bosons.}

Since $r$ parametrizes how much the gauge symmetry of $Z_L$ is spontaneously broken, quantum radiative corrections to $m_{Z_L}^2$ would vanish in the $r \to 0$ limit to enhance the gauge symmetry. 
In this sense, a small $m_{Z_L}^2$ is technically natural \cite{'tHooft:1979bh}. 
While any spin objects (scalar, fermion, vector, etc.) with the same mass texture should give the same results\footnote{See Ref.~\cite{Kim:2004rp} for the natural inflation with multi-axion, where specific alignment of couplings of axions to non-Abelian instantons allows a flat direction, along which an effective axion decay constant can be enhanced.}, it is a superior part of the vector gauge boson case that its gauge symmetry will automatically protect the small mass from the loop corrections.

The gauge see-saw mechanism relies on the large mixing among the interaction eigenstates.
In the perfect mass alignment case (with a zero eigenvalue), the mixing angle is given by
\bea
\sin\theta=\sqrt{\frac{a}{a + d}} \, , \quad
\cos\theta=\sqrt{\frac{d}{a + d}} \, .
\eea
The texture in Eq.~\eqref{eq:M1111} would give the maximal mixing ($\theta = \pi / 4$) of this case.

%%%%%%%%%%%%%%%%%%%%%%%%%%%%
\subsubsection{Illustrations}
%%%%%%%%%%%%%%%%%%%%%%%%%%%%
The gauge see-saw can work for any model that gives the masses to two $U(1)$s simultaneously.
It can be extended to a larger number of the $U(1)$s in a straightforward way.
We illustrate the realization of the gauge see-saw in the mass matrix using the Higgs mechanism and the St\"uckelberg mechanism.

We take two Abelian gauge groups: $U(1)'$ with a gauge boson $\hat Z'$ and a gauge coupling constant $g'$, and $U(1)''$ with $\hat Z''$ and $g''$.

\vspace{3mm}
\noindent
(i) Using Higgs mechanism: \\
In this realization, we first assume the couplings of the $\hat Z'$, $\hat Z''$ to the SM fermions are vectorial.
Otherwise, the SM Higgs contribution to the mass matrix should be considered, which is beyond the scope of our simple illustration.

We consider two SM singlet complex scalars to break the two gauge symmetries spontaneously: $\Phi_1$ with a $U(1)'$ charge $q'_1$, a $U(1)''$ charge $q''_1$, a vev $v_1$, and $\Phi_2$ with $q'_2$, $q''_2$, $v_2$.
The relevant Lagrangian is given by
\bea
{\cal L} \sim \sum_{i=1,2} \left| \big(\partial_\mu + i g' q'_i \hat{Z}'_\mu + i g'' q''_i \hat{Z}''_\mu \big) \Phi_i \right|^2 .
\eea
The mass-squared matrix for the gauge bosons in the $(\hat Z', \hat Z'')$ basis is given by
\bea
M = \mat{g'^2 (q'^2_1 v_1^2 + q'^2_2 v_2^2) & g' g'' (q'_1 q''_1 v_1^2 + q'_2 q''_2 v_2^2) \\
g' g'' (q'_1 q''_1 v_1^2 + q'_2 q''_2 v_2^2) & g''^2 (q''^2_1 v_1^2 + q''^2_2 v_2^2)} . \nn \label{eq:Mhiggs}  \\ 
\eea
Then $\dM = g'^2 g''^2 ( q'_1 q''_2 - q''_1 q'_2 )^2 v_1^2 v_2^2$, which tells the perfect mass alignment case is achieved for $q'_1 q''_2 - q''_1 q'_2 = 0$.

For $(q'_1 q''_2 - q''_1 q'_2)^2 \ll 1$, the gauge see-saw mechanism works ($r \ll 1$), and the physical masses are approximated by
\bea
m_{Z_L}^2 &\approx& \frac{g'^2 g''^2 (q'_1 q''_2 - q''_1 q'_2)^2 v_1^2 v_2^2}{(g'^2 + g''^2 (q''^2_2 / q'^2_2)) (q'^2_1 v_1^2 + q'^2_2 v_2^2)} , \\
m_{Z_H}^2 &\approx& (g'^2 + g''^2 (q''^2_2 / q'^2_2)) (q'^2_1 v_1^2 + q'^2_2 v_2^2) .
\eea

In the case of $g' \sim g''$, $v_1 \sim v_2$, $q'_1 \sim q''_1 \sim q''_2 \sim q'_2 \sim {\cal O}(1)$, we get
\bea
m_{Z_L}^2 &\sim& {\cal O}(1) \, g'^2 v_1^2 (q'_1 q''_2 - q''_1 q'_2)^2 , \label{eq:mZLapprox} \\ 
m_{Z_H}^2 &\sim& {\cal O}(1) \, g'^2 v_1^2 , \label{eq:mZHapprox}
\eea
which clearly shows that $m_{Z_H}$ stays at the original scale while $m_{Z_L}$ is suppressed by the small mass differences (or charge differences) in Eq.~\eqref{eq:Mhiggs}, giving $r \sim {\cal O}(1) \, (q'_1 q''_2 - q''_1 q'_2)^2$.

If the two $U(1)$s are re-defined to have only diagonal masses $(m_{Z_L}^2 , \, m_{Z_H}^2)$, then the two Higgs scalars become linear combinations of each other with mixed $U(1)$ charges and vevs.
One can see the gauge see-saw mechanism works only when one of these combinations has small mixed $U(1)$ charges and vevs.

\vspace{3mm}
\noindent 
(ii) Using St\"uckelberg mechanism: \\
In the St\"uckelberg mechanism \cite{Stueckelberg:1900zz,Kors:2004dx,Feldman:2007wj}, we do not need real scalars, but need at least two pseudoscalars ($a_1$, $a_2$) transforming non-linearly under the two $U(1)$s.

Under the $U(1)'$, they transform as
\bea
&&a_1 \to a_1 - c'_1 \lambda'(x) , \quad a_2 \to a_2 - c'_2 \lambda' (x) , \\
&&\qquad ~~\text{while}~~ \hat{Z}'_\mu \to \hat{Z}'_\mu +\partial_\mu \lambda' (x) ,
\eea
and similarly for the $U(1)''$.

%In some sense, the St\"uckelberg mechanism can be viewed as a decoupling limit of the Higgs mechanism case, where Higgs bosons are integrated out. 
%The coefficients $c'$, $c''$ can be considered as the $U(1)$ coupling and charge for the Higgs scalars in this picture.

With two gauge invariant combinations $\partial_\mu a_1 +c'_1 \hat{Z}'_\mu + c''_1 \hat{Z}''_\mu$ and $\partial_\mu a_2 +c'_2 \hat{Z}'_\mu + c''_2 \hat{Z}''_\mu$, the mass terms are given by
\bea
{\cal L} \sim \sum_{i=1,2} \frac12 \rho_i^2 \big( \partial_\mu a_i +c'_i \hat{Z}'_\mu + c''_i \hat{Z}''_\mu \big)^2 ,
\eea
with some mass parameters $\rho_1$ and $\rho_2$,
giving the mass-squared matrix
\bea
M = \mat{ c'^2_1 \rho_1^2 + c'^2_2 \rho_2^2 & c'_1 c''_1 \rho_1^2 + c'_2 c''_2 \rho_2^2 \\
c'_1 c''_1 \rho_1^2 + c'_2 c''_2 \rho_2^2 & c''^2_1 \rho_1^2 + c''^2_2 \rho_2^2}. ~~~~
\eea
The choice of coefficients satisfying $(c'_1 c''_2 - c''_1 c'_2)^2 \ll 1$ turns on the gauge see-saw condition, which makes one gauge boson much lighter than the other.
The remaining part resembles the Higgs mechanism case.

\vspace{3mm}
We emphasize that the modeling through the gauge see-saw mechanism may not be particularly natural compared to other options, such as taking a small gauge coupling.
Rather, what this mechanism suggests is there is another way to see the origin of a light gauge boson.
Although it might be a drawback to introduce a certain fine-tuning among the charges, it is possible to find a UV origin where only higher scale vev's with no small quantity is introduced in order to have a light gauge boson.

%%%%%%%%%%%%%%%%%%%%%%%%%%%%
\subsubsection{Types of the $U(1)$ symmetries}
%%%%%%%%%%%%%%%%%%%%%%%%%%%%
Now, we want to consider the constraints on the type of the $U(1)$ gauge symmetries.

The gauge see-saw mechanism works for any two disparate scales, for instance, two $U(1)$s originally at the GUT scales and the TeV scale $Z_L$ as a result of the gauge see-saw.\footnote{In this regard, it is interesting to note that the $E_6$ grand unified theories can provide two $U(1)$s, $E_6 \to U(1)_\psi \times U(1)_\chi \times SU(5)$.}
However, in this letter we emphasize the case where the light gauge boson is the sub-electroweak scale (such as the dark photon \cite{ArkaniHamed:2008qn} and the dark $Z$ \cite{Davoudiasl:2012ag}).
For such a light new particle to survive all the experimental constraints, its coupling to the SM particles should be very small.

Although it is possible to impose a tiny gauge coupling to avoid experimental constraints, it would also bring down the gauge boson mass, which would be out of our spirit of using the gauge see-saw to explain the light gauge boson.
Then we are left with two options.
One is taking the cancellation of the two gauge couplings in the physical eigenstates.
The other is taking the $U(1)$s as dark gauge symmetries under which the SM particles do not have charges, and the gauge bosons interact with the SM particles only through small mixing.
We will consider each case one by one.

\vspace{3mm}
\noindent 
(i) Twin gauge symmetries: \\
The vector couplings of matter currents to gauge bosons can be written as
\bea
{\cal L}_\text{int} & \sim & - \big[\cos\theta (g' J'_\mu) -\sin\theta (g'' J''_\mu)\big] Z_L^\mu \nn \\
& & - \big[\sin\theta (g' J'_\mu) + \cos\theta (g'' J''_\mu)\big] Z_H^\mu, \label{eq:interaction}
\eea
in terms of the gauge boson mass eigenstates.

For the sub-electroweak scale $Z_L$ to survive the experimental constraints \cite{Batell:2014yra,Heeck:2014zfa,Jeong:2015bbi,Altmannshofer:2016brv}, its coupling should be greatly suppressed ($\cos\theta g' J'_\mu \simeq \sin\theta g''J''_\mu$).
This leads us to contemplate the two $U(1)$ gauge symmetries are twin.

For an example, we can consider two $U(1)_{B-L}$ symmetries so that $J'_\mu = J''_\mu = J^{B-L}_\mu = \sum_i  (B-L)_i \bar{f}_i \gamma_\mu f_i$ with $g' \simeq g''$.
Then the $Z_L$ coupling is greatly suppressed near the maximal mixing limit of Eq.~\eqref{eq:M1111}, i.e., $\theta \simeq \pi/4$.
In this case, Eq.~\eqref{eq:interaction} can be well approximated by 
\bea 
{\cal L}_\text{int} \sim - J^{B-L}_\mu \big[ (\cos\theta g' -\sin\theta g'') Z_L^\mu + \sqrt{2} g' Z_H^\mu \big] , \nn \\
\eea
where the $Z_L$ is a light $B-L$ gauge boson whose coupling is suppressed by the cancellation, while the $Z_H$ is essentially the typical heavy $B-L$ gauge boson with some enhancement in its coupling.
 
Such a light $B-L$ gauge boson $Z_L$ with a suppressed coupling can be used in many models including the freeze-in right-handed neutrino dark matter scenario \cite{Kaneta:2016vkq}.
 
\vspace{3mm}
\noindent 
(ii) Kinetic mixings: \\
In general, there are kinetic mixings among the $U(1)$ gauge symmetries \cite{Holdom:1985ag}.
The kinetic mixing between the dark $U(1)$ and the $U(1)_Y$  can provide small coupling for a new gauge boson to the electromagnetic current ($J_\text{EM}$) and the weak neutral current ($J_\text{NC}$) even in the absence of direct couplings to the SM particles.

We may consider two kinetic mixings: $U(1)'-U(1)_Y$ mixing and $U(1)''-U(1)_Y$ mixing.
\bea
{\cal L}_\text{kin mix} \sim \frac{\varepsilon_1}{2\cos\theta_W}\hat{Z}'_{\mu\nu}B^{\mu\nu} + \frac{\varepsilon_2}{2\cos\theta_W}\hat{Z}''_{\mu\nu}B^{\mu\nu} . \label{eq:kineticmixings}
\eea
A $U(1)'-U(1)''$ mixing ($\frac{\varepsilon_{12}}{2} \hat{Z}'_{\mu\nu}\hat{Z}''^{\mu\nu}$) would not affect any physics unless other dark sector particles such as the dark matter are introduced.

In terms of the effective kinetic mixing parameters for the mass eigenstates
\bea
\varepsilon_L = \cos\theta\varepsilon_1-\sin\theta\varepsilon_2 , ~~ \varepsilon_H = \sin\theta \varepsilon_1 +\cos\theta\varepsilon_2 ,
\eea  
one can obtain the interaction Lagrangian for the mass eigenstates $Z_L$ and $Z_H$ in the leading order of the $\varepsilon_i$ as
\bea
{\cal L}_{Z_L} &\sim& - \varepsilon_L \Big[ e J^\text{EM}_\mu + \frac{m_{Z_L}^2}{m_{Z}^2} t_W g_Z J^\text{NC}_\mu \Big] Z_L^\mu , \\
{\cal L}_{Z_H} &\sim& - \varepsilon_H \Big[ e J^\text{EM}_\mu + \frac{m_{Z_H}^2}{m_{Z}^2 - m_{Z_H}^2} t_W g_Z J^\text{NC}_\mu \Big] Z_H^\mu ,
\eea
where $t_W \equiv \tan\theta_W$ is the weak mixing angle factor, and $g_Z = g / \cos\theta_W$.
The kinetic mixing case can be also viewed as a kind of twin symmetry as far as the SM sector is concerned.

We took the $m_{Z_L} \ll m_Z$ limit, which has a suppression of the $m_{Z_L}^2 / m_Z^2$ in the coupling to the weak neutral current.
This is consistent with that the massless gauge boson cannot have an axial coupling \cite{Lee:2016ief}.

The constraints for the kinetic mixing parameter can be found in Refs.~\cite{Essig:2013lka,Hook:2010tw,Batley:2015lha}.

%%%%%%%%%%%%%%%%%%%%%%%%%%%%
\subsubsection{Discussions}
%%%%%%%%%%%%%%%%%%%%%%%%%%%%
The gauge see-saw mechanism not only provides an explanation of the mass of a light gauge boson $Z_L$, but also implies a rich phenomenology partly because of its connection to a heavy gauge boson $Z_H$.
We discuss some of them here very briefly.

If the $Z_L$ is a sub-electroweak scale gauge boson, which is severely constrained, both $Z_L$ and $Z_H$ are necessarily the same kind (twin symmetries or kinetic mixings) except for the overall strength.
It can allow sufficiently suppressed coupling for the $Z_L$.
Measurement and comparison of the couplings of two gauge bosons, if discovered, can be an important test of whether the gauge see-saw mechanism is underneath the two disparate scale gauge bosons.

In the scenario of the twin $B-L$, although the light $Z_L$ coupling to the $J_{B-L}$ is suppressed, the heavy $Z_H$ coupling to the same current is not suppressed.
Thus the heavy TeV-scale gauge boson signal can be large enough to be observed at the typical heavy resonance searches at the LHC experiments \cite{CMS:2015nhc,Aaboud:2016cth}, while the light gauge boson can avoid severe constraints \cite{Heeck:2014zfa}.

For the kinetic mixing scenario, the heavy $Z_H$ has a rather sizable coupling to the weak neutral current $J_\text{NC}$ while the light $Z_L$ has a suppressed coupling (roughly, $\propto m_{Z_L}^2 / m_Z^2$) to the $J_\text{NC}$.
This is a character of a typical dark photon, which means typical dark photon scenario can be realized as a part of the two dark $U(1)$s with a gauge see-saw.
(For discussion on this suppression, see Ref.~\cite{Davoudiasl:2012ag} for example.)

It would be interesting to consider some physics cases where both $Z_H$ and $Z_L$ appear.
There are already some studies of simultaneous use of a heavy and a light gauge boson in the literature.
For instance, in Ref.~\cite{Buschmann:2015awa}, a TeV-scale heavy gauge boson is produced by the Drell-Yan process at the LHC and decays into a pair of the dark matter particles, which subsequently radiates off new light GeV-scale gauge bosons.
This kind of scenario can be naturally implemented using the gauge see-saw.

Some on-shell channels such as $Z_H \to Z_L + Z$, $Z_H \to Z_L + Z_L$, and $Z_H \to 3 Z_L$ would be also possible depending on the mediator particles.
For instance, the last case in which a heavy $Z_H$ decaying into 3 pairs of the dilepton ($Z_L \to \ell^+\ell^-$), through the scalar mediators and a large mixing between $\hat Z'$ and $\hat Z''$, making a resonance at the $Z_H$ (in a somewhat similar fashion to Ref.~\cite{Barger:2009xg}) would be an interesting signal.

Plenty more implications of the gauge see-saw in collider physics and dark matter physics are warranted.

%%%%%%%%%%%%%%%%%%%%%%%%%%%%
\subsubsection{Summary and Outlook}
%%%%%%%%%%%%%%%%%%%%%%%%%%%%
A heavy gauge boson search at the energy frontier (using Drell-Yan process at the high-energy colliders) and a very light gauge boson search at the intensity frontier (using fixed-targets or meson decays) are two popular direct bump searches of a new gauge interaction.
While one search may not have much to do with the other, it is interesting to note a possible connection of the two through the mass mixing.

We presented a mechanism that can explain how a dark gauge boson scale can be so small compared to the usual new physics scale, TeV or UV scale.
In its essence, the gauge see-saw mechanism is a way to shift a question of why (physical eigenstate) mass difference is so huge ($m_{Z_L} \ll m_{Z_H}$) into a question of why (interaction eigenstate) mass difference is so tiny ($\dM \ll \tM^2$).

Like the neutrino see-saw mechanism, the gauge see-saw mechanism is found to have substantial potential of applications in modeling and phenomenology in broad areas of particle physics in low-energy, high-energy and cosmology, as some of them were briefly mentioned in this letter.

%---------------------------------------------------------
\vspace{3mm}
\begin{acknowledgments}
Acknowledgments: This work was supported by IBS (Project Code IBS-R018-D1).
We thank E. Ma for very helpful discussions about the gauge see-saw mechanism during the Light Dark World International Forum 2016.
HL thanks H. Davoudiasl and W. Marciano for a long-term collaboration on a light gauge boson.
MS thanks DY Kim for a useful discussion.
\end{acknowledgments}
%---------------------------------------------------------

%---------------------------------------------------------

\end{document}